\documentstyle[prd,aps,floats,tabularx,epsfig]{revtex}

\begin{document}
\draft
\preprint{\
\begin{tabular}{rr}
&
\end{tabular}
}
\twocolumn[\hsize\textwidth\columnwidth\hsize\csname@twocolumnfalse\endcsname
\title{Have Acoustic Oscillations been Detected in the Current Cosmic Microwave
Background Data?}
\author{M.~Douspis \&
  P.~G.~Ferreira}
\address{Astrophysics, University of Oxford,
NAPL, Keble Road, Oxford OX1 3RH, UK}

\maketitle

\begin{abstract}
The angular power spectrum of the Cosmic Microwave Background has
been measured out to sufficiently small angular scale to encompass
a few acoustic oscillations. We use a phenomenological fit to the
angular power spectrum to quantify the statistical significance 
of these oscillations and discuss the cosmological implications of 
such a finding.
\end{abstract}
\date{\today}
]
\renewcommand{\thefootnote}{\arabic{footnote}} \setcounter{footnote}{0}
\noindent
During the past few years there has been unprecedented progress in
the detection and characterization of fluctuations in the Cosmic
Microwave Background on small angular scales. The TOCO, BOOMERanG
and MAXIMA experiments \cite{3exp} presented unambiguous evidence for a preferred
scale (a ``peak'') in the variance of temperature fluctuations
as a function of scale,
$(\Delta T)^2_\ell\equiv \ell(\ell+1)C_\ell/4\pi$ (where $C_\ell$ is the
angular power spectrum at a scale $\ell\simeq 180^{\rm 0}/\theta$). This is
a striking result, having been predicted in 1970 from simple
assumptions about scale-invariance and linear perturbation
theory of General Relativity.  During the last six months
the possible presence of peaks and troughs in $(\Delta T)^2_\ell$
has been reported by the BOOMERanG \cite{BOOM}, DASI 
\cite{DASI} and MAXIMA \cite{MAXIMA} experiments.
This harmonic set of features in the power spectrum could
further confirm that the origin of structure was due to a
primordial set of fluctuations which set the cosmological
plasma ``ringing'' at very early times. It is the purpose of
this short note to quantify the confidence with which one
can claim that there are acoustic oscillations present
in the current data. 

We shall first outline the argument for why we expect oscillations
in the power spectrum of the CMB. Let us restrict ourselves to
the radiation era, where the photons and baryons are tightly coupled.
Its suffices to consider the density contrast in radiation, $\delta_\gamma$
and (in the synchronous gauge) the trace of the spatial metric
perturbations $h\equiv h^i_i$. The first order Einstein and continuity
equations in Fourier space are:
\begin{equation}
{\ddot h}+\frac{1}{\eta}{\dot h}+\frac{6}{\eta^2}\delta_\gamma=0 \ \ \mbox{\rm and} \ \ 
{\ddot \delta_\gamma}+\frac{k^2}{3}\delta_\gamma+\frac{2}{3}{\ddot h}=0
\label{eveq} 
\end{equation}
where $\eta$ is conformal time and $k$ labels the Fourier
component. Let us 
restrict ourselves to adiabatic initial conditions. In this
case one has that $\delta_\gamma=-\frac{2}{3}h$. One can solve 
Eq. \ref{eveq} on large scales $k\eta\ll1$ to find two solutions
$\delta_\gamma=A_{\bf k}\eta^2,B_{\bf k}\eta^{-2}$; if these are setup
early in the radiation era, the growing dominates very rapidly
and one can to as excellent approximation set $B_{\bf k}=0$. 
On small scales one can solve the system using a WKB approximation
to find $\delta_\gamma \propto \cos(k\eta/\sqrt{3}), \sin(k\eta/\sqrt{3})$.
Matching the large scale solution to the small scale solution
one finds that $\delta_\gamma=A_{\bf k}(k\eta)^2\cos(k\eta/\sqrt{3}+\phi)$. 
The acoustic oscillations in the $\Delta T^2_\ell$
will be primarily the power spectrum of $\delta_\gamma$ at recombination,
$\eta_*$ projected until today:
\begin{equation}
\Delta T^2_\ell\propto \langle|A_{\bf k}|^2\rangle \cos^2(\frac{k\eta_*}{\sqrt{3}}+\phi)\mid_{\ell=k\chi(\eta_0-\eta_*)}
\end{equation}
where $\chi(r)$ is the conformal distance corresponding to the coordinate
distance $r$. Thus, the oscillations in the baryon-photon
plasma will lead to a set of peaks and troughs in the angular power spectrum.

\begin{figure}[!t]
\begin{center}
\includegraphics[angle=0,totalheight=7cm,width=8.cm]{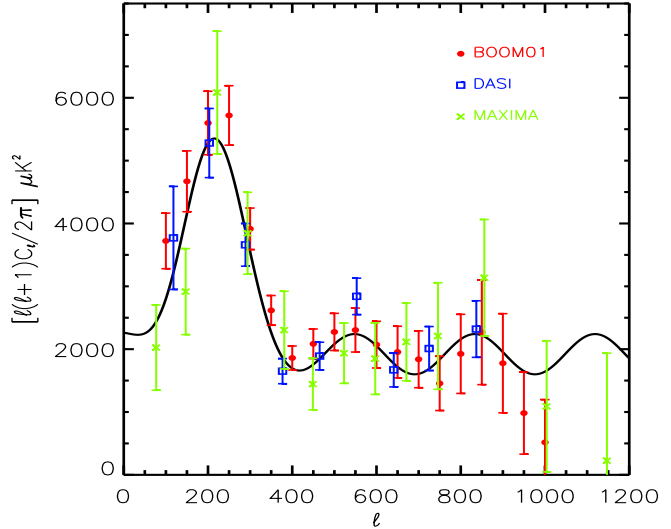}
\end{center}
\caption{\label{plotFBP}Data from BOOMERANG, DASI and MAXIMA. Overplotted, our best fit from the class of models described by Eq. 3.}
\end{figure}

How general is this argument? As stated above we are considering
 primordial, passive, perturbations which are in effect initial
conditions for the evolution of the coupled baryon/photon plasma
responding to gravity and pressure \cite{decoh}. The most general class
of such perturbations was classified in \cite{BMT} where, in the
context of the current menagerie of matter candidates, it is believed 
that there are five degrees of freedom, possibly correlated among
each other (corresponding to a $5\times5$ symmetric matrix of initial
conditions). 
Although we looked at the specific case of adiabatic perturbations, the
argument follows through for all other {\it pure} primordial perturbations.
By pure perturbations we means perturbations in which one only picks
one of these degrees of freedom to be non-zero.
The key feature is that the large scale solutions to Eq. \ref{eveq}
have two modes, one of which is decaying and very rapidly becomes 
subdominant. For example for density isocurvature perturbations
a different phase, $\phi$ will be picked out such that
the positions of the peaks will
be out of phase with regards to the adiabatic models.  If one considers
 sums of initial condition such as in \cite{BMT}, then
it is conceivable that the combination of power spectra will interfere
in such a way as washout the oscillations. 

Clearly, the presence of oscillatory features would be strong evidence
that the structure was seeded at some early time, before recombination.
With the current CMB data it has been suggested that we are already
seeing evidence for such features. One of the reasons for such a claim
is that primordial, passive models supply a good fit to the measured
angular power spectrum and, as argued above these models have acoustic
oscillations. The concern is that, {\it all} models which have been
compared to the data have oscillations in the $C_\ell$ and one therefore
has not strictly tested for the presence of these oscillations. We
propose to do this in the following: construct a phenomenological fit 
to the data points that can smoothly interpolate between a model
with no oscillations to one which has oscillations of a well defined
frequency and phase. An analogous approach was used in \cite{KP}
to quantify the significance of the presence of a peak at $\ell\simeq200$.
The parametrization we shall use is of the following form:
\begin{equation}
\Delta T^2_\ell=Ae^{-\frac{(\ell-\ell_{p})^2}{2\sigma^2}} 
+B\cos^2[\alpha(\ell-\ell_{p})+\phi]+C \label{eqmod}
\end{equation}
This is a seven parameter family of models and we can justify it as 
follows. The goal is to detect the presence of prefered frequency
in the $\Delta T^2_\ell$ so the parameters we are ultimately interested
in are $\alpha$ and $B$. However we know that there is a well defined
peak in the data with a well defined width, and the signal to noise
of this peak is sufficiently high that it will dominate any statistical
analysis; i.e. the width of the peak, $\sigma$ will tend to peak at
a frequency of order $\pi/\sigma$. Given that we wish to be conservative
we consider a part of the curve corresponding to the first peak 
(characterized by $A$, $\ell_{p}$ and $\sigma$) and marginalize over these
parameters. Hopefully in this way we decrease the statistical weight of
the peak. Finally we introduce an offset, $C$ which allows us to interpolate
between a flat and oscillatory curve.

Given that we are interested in the behaviour of the power spectrum in the
regime where acoustic oscillations will dominate, i.e. on scales larger
than the sound horizon at last scattering, we shall not include the
COBE data set. We restrict ourselves to three data sets, BOOMERanG \cite{BOOM},
DASI \cite{DASI} and MAXIMA \cite{MAXIMA}. We shall minimize the fitting
function of Eq. \ref{eqmod} with regards to the three data sets using
a standard $\chi^2$. A few comments are in order. We do not consider 
more refined parametrizations of the band-power distribution functions\cite{BJK}, such as the off-set log-normal or the skewed approximation. 
We marginalize over the calibration uncertainties quoted in \cite{BOOM,DASI,MAXIMA}. Furthermore the beam uncertainties are taken into account by
adding them in quadrature to the noise covariance matrices of each experiment.
All these approximations may introduce a modest degree of uncertainty in
our results but do not change the essential conclusions.

\begin{table}
\begin{center}
\begin{tabular}{|l|c|c|c|}
\hline Experiment & $\alpha\times10^{-2}$ &
 $B\times10^3$ ($\mu$K)$^{2}$ & $\chi^2$ ($N_D$)\\
\hline  
All & $1.1^{+0.2}_{-0.4}$&$0.7^{+0.3}_{-0.3}$   & $30$ ($42$) \\
All$_{>400}$ & $1.0^{+0.3}_{-0.6}$ &$0.8^{+0.3}_{-0.3}$ & $16$ ($27$) \\
BOOMERanG & $0.8^{+0.6}_{-0.7}$ & $0.7^{+0.4}_{-0.4}$ &$6$ ($19$) \\
MAXIMA & $0.7^{+0.8}_{-0.6}$ & $1.0^{+1.0}_{-0.9}$ &$3$ ($14$)\\
DASI & $1.1^{+0.3}_{-0.3}$ & $1.1^{+0.4}_{-0.5}$ &$2$ ($9$)\\
DASI$_{-1}$ & $0.9^{+0.6}_{-0.8}$ & $0.9^{+1.0}_{-0.7}$ &$1$ ($8$)\\
MOCK & $1.0^{+0.3}_{-0.3}$ & $0.7^{+0.3}_{-0.3}$ &$34$ ($42$)\\
MOCK$_{>400}$ & $1.0^{+0.3}_{-0.3}$ &  $0.6^{+0.3}_{-0.3}$ &$27$ ($27$)\\
\hline
\end{tabular}
\end{center}
\caption{The mean values and 95\% errors for the $\alpha$ and $B$ as defined
in Equation 3 and the corresponding $\chi^2$ for the best fit models (where
$N_D$ is the number of parameters). See text for description of the
different data combinations.}
\label{table1}
\vskip -.2in
\end{table}

In table \ref{table1} we summarize our results for $\alpha$, $B$ and the
$\chi^2$ of the best fit model to each subset of data. The different
combination and subsets of the data we consider are: data from all
three experiments (``All'') and for all three data experiments
discarding all points with $\ell<400$ (``All$_{>400}$''), the data from each
individual experiment (``BOOMERanG'', ``MAXIMA'' and ``DASI''), data
from DASI discarding the point at $\ell=553$ (``DASI$_{-1}$'').
We have also generated a mock realization of the 
best fit adiabatic model to the data with
corresponding variance from sampling and noise from the combination of
the three experiments (``Mock'') and finally the same realization but discarding
all points with $\ell<400$ (``Mock$_{>400}$'').
We present the mean values and the $95\%$ confidence limits from the
integrated likelihoods. We find that the
data does seem to pick out a favorite frequency of oscillation
in the $\Delta T^2_\ell$ of $\alpha=1.1^{0.2}_{-0.1}$,
 corresponding to an interpeak spacing
of $\delta\ell=\pi/\alpha=286^{+163}_{-44}$. It is interesting
to note that even removing the points that lie in the region
of the 1$^{\rm st}$ peak, the detection persists albeit with larger
error bars. The three experiments detect similar values of $\alpha$ with
varying confidence regions. We should note that the likelihoods
in $\alpha$ are extremely skewed and in some cases actually have two
local maxima. For example the maximum of the likelihood for BOOMERanG
is $\alpha=0.01$ while for MAXIMA there is a local maximum at 
$\alpha=0.011$. Note also the importance of the point at $\ell=553$
in the DASI data. If we remove this point from our analysis, the
DASI confidence region for $\alpha$ enlarges considerably. 
We have listed the values of the $\chi^2$ for the corresponding
best fit models along with the number of data points used in
each case. All of the $\chi^2$ are reasonable although 
correlations between the data points may lead to a smaller number
of degrees of freedom than simply $N_D-N_P$ where $N_P=6$ is
the number of parameters. The Mock data sets lead to similar
values of $\alpha$ and corresponding confidence limits.

Our analysis indicates that there is a marginal presence of
oscillations in the measured $\Delta T^2_\ell$ (at the $2-\sigma$
level) {\it within the context of the phenomenological model described
by Eq. 3}.
This caveat is important. Although we have attempted to justify
the functional form of our model using rough general arguments,
it is conceivable that one could construct other models which interpolate
between oscillatory and non-oscillatory behaviour and which 
reduce or enhance the significance of detection of $\alpha$. For
example, if we consider the All$_{>400}$ combination of data
and add a term of the form  $D\ell$ to Equation 3., the mean value of
$\alpha$ is still $0.01$ but now $\alpha=0$  lies within
the $95\%$ confidence region. We have, of course, done this by adding yet
another parameter. However this does not exclude the possibility
that there are models with fewer parameters and no oscillations that may
provide a better fit to the data.

Let us pursue the implications of the constrain on $\alpha$
within the context of models
which predict oscillations. The above discussion leads us to note that
the spacing between peaks $\delta\ell$ can be used to set a constraint
on the angular-diameter distance. As noted in \cite{HW} the physical peak
separation is set by the sound-horizon (which we know from atomic physics).
Combined with the our knowledge of time of recombination and $\delta\ell$
(which is effectively half the angular scale subtended by the sound
horizon at recombination) we can constrain the geometry of the universe. 
Efstathiou and Bond \cite{EB} have proposed a convenient parametrization
using the ``shift'' parameter $R=2\sqrt{\Omega_K/\Omega_M}/\chi(\eta_0-\eta_*)$
where $\Omega_K=1-\Omega_\Lambda-\Omega_M$ and $\Omega_M$ ($\Omega_\Lambda$)
are the fractional energy densities of matter (cosmological constant) today.
Indeed one has that $\delta\ell\simeq290/R$ and one can obtain a likelihood
for $R$: one finds that $R=1.01^{+0.18}_{-0.37}$. This can be
reexpressed into a constrain on $\Omega=1-\Omega_K$ if we marginalize
over $\Omega_\Lambda$ to obtain $\Omega \simeq 1^{+0.24}_{-0.18}$. Note that
this constraint is {\it independent} from the constraint obtained
from the assumption of adiabaticity and the position of the first
peak, $\ell_p\simeq215/R$ where one finds $R=1.05^{+0.05}_{-0.04}$
and $\Omega \simeq 1^{+0.2}_{-0.12}$. These
two independent constraints on $R$ and $\Omega$ are consistent.

In summary we have attempted to assess the significance of the presence
of acoustic oscillations in current CMB data using a model independent
method. One could view Eq. \ref{eqmod} as a different class of phenomenological
models which has the advantage (given the question we are attempting to
answer) of smoothly interpolating between the absence and presence of
acoustic oscillations. We subsequently use the interpeak separation
(or acoustic oscillation frequency) to derive a constraint on the 
geometry of the universe. Such a constraint is independent to that
derived from the position of the first peak. However it is necessarily
included in any analysis which considers the full set of initial
conditions put forward in \cite{BMT}. With the rapid progress of
CMB experiments, such an analysis will become essential to obtain
accurate, model independent constraints on the cosmological parameters.

{\it Acknowledgments}: 
PGF acknowledges support from the Royal Society. The authors thank F. Vernizzi, M. Kunz, A. Melchiorri and the MAXIMA collaboration
 for discussions and advice.

\tighten

\end{document}